\documentclass[aps,prl,showpacs,twocolumn,superscriptaddress,floatfix]{revtex4-2}
\usepackage[utf8]{inputenc}
\newcommand{\figurescale}{1}
\usepackage{verbatim}
\usepackage{amsmath}
\usepackage{mathtools}

\DeclarePairedDelimiterX\braket[2]{\langle}{\rangle}{#1 \delimsize\vert #2}

\usepackage[pdftex]{graphicx}
\usepackage[separate-uncertainty=true]{siunitx}
\DeclareSIUnit{\rpm}{rpm}
\usepackage[colorlinks=true,urlcolor=blue,linkcolor=blue,citecolor=blue]{hyperref}
\usepackage[usenames,dvipsnames]{xcolor}
\usepackage[T1]{fontenc}
\usepackage{placeins}
\usepackage{nicefrac}
\usepackage{braket}
\usepackage{pdfcomment}
\usepackage{xcolor}
\usepackage{braket}
\usepackage[normalem]{ulem}

\usepackage{lineno}

\begin{document}

\title{Controlling exciton many-body states by the electric-field effect in monolayer MoS$_2$}
%
\author{J.~Klein}\email{jpklein@mit.edu}
\affiliation{Walter Schottky Institut and Physik Department, Technische Universit\"at M\"unchen, Am Coulombwall 4, 85748 Garching, Germany}
\affiliation{Department of Materials Science and Engineering, Massachusetts Institute of Technology, Cambridge, Massachusetts 02139, USA}
\author{A.~H\"otger}
\affiliation{Walter Schottky Institut and Physik Department, Technische Universit\"at M\"unchen, Am Coulombwall 4, 85748 Garching, Germany}
\author{M.~Florian}
\affiliation{Institut für Theoretische Physik, Universität Bremen, P.O. Box 330 440, 28334 Bremen, Germany}
\author{A.~Steinhoff}
\affiliation{Institut für Theoretische Physik, Universität Bremen, P.O. Box 330 440, 28334 Bremen, Germany}
\author{A.~Delhomme}
\affiliation{Universit\'e Grenoble Alpes, INSA Toulouse, Univ. Toulouse Paul Sabatier, EMFL, CNRS, LNCMI, 38000 Grenoble, France.}
\author{T.~Taniguchi}
\affiliation{Research Center for Functional Materials, National Institute for Materials Science, 1-1 Namiki, Tsukuba 305-0044, Japan}
\author{K.~Watanabe}
\affiliation{Research Center for Functional Materials, National Institute for Materials Science, 1-1 Namiki, Tsukuba 305-0044, Japan}
\author{F.~Jahnke}
\affiliation{Institut für Theoretische Physik, Universität Bremen, P.O. Box 330 440, 28334 Bremen, Germany}
\author{A.~W.~Holleitner}
\affiliation{Walter Schottky Institut and Physik Department, Technische Universit\"at M\"unchen, Am Coulombwall 4, 85748 Garching, Germany}
\author{M.~Potemski}
\affiliation{Universit\'e Grenoble Alpes, INSA Toulouse, Univ. Toulouse Paul Sabatier, EMFL, CNRS, LNCMI, 38000 Grenoble, France.}
\author{C.~Faugeras}
\affiliation{Universit\'e Grenoble Alpes, INSA Toulouse, Univ. Toulouse Paul Sabatier, EMFL, CNRS, LNCMI, 38000 Grenoble, France.}
\author{J.~J.~Finley}\email{finley@wsi.tum.de}
\affiliation{Walter Schottky Institut and Physik Department, Technische Universit\"at M\"unchen, Am Coulombwall 4, 85748 Garching, Germany}
\author{A.~V.~Stier}\email{andreas.stier@wsi.tum.de}
\affiliation{Walter Schottky Institut and Physik Department, Technische Universit\"at M\"unchen, Am Coulombwall 4, 85748 Garching, Germany}
%
%
\date{\today}
%
%
\begin{abstract}
We report magneto-optical spectroscopy of gated monolayer MoS$_2$ in high magnetic fields up to $\SI{28}{\tesla}$ and obtain new insights on the many-body interaction of neutral and charged excitons with the resident charges of distinct spin and valley texture. For neutral excitons at low electron doping, we observe a nonlinear valley Zeeman shift due to dipolar spin-interactions that depends sensitively on the local carrier concentration. As the Fermi energy increases to dominate over the other relevant energy scales in the system, the magneto-optical response depends on the occupation of the fully spin-polarized Landau levels in both $K/K^{\prime}$ valleys. This manifests itself in a many-body state. Our experiments demonstrate that the exciton in monolayer semiconductors is only a single particle boson close to charge neutrality. We find that away from charge neutrality it smoothly transitions into polaronic states with a distinct spin-valley flavour that is defined by the Landau level quantized spin and valley texture.
\end{abstract}
%
%
\maketitle
%
%

A bosonic exciton immersed in a Fermi gas is an example of the \emph{impurity-model} that can be understood in terms of a polaronic picture - dressing of the exciton by collective excitations of the many-body environment~\cite{Massignan.2014,Bloch.2008,Boll.2016,Mazurenko.2017,Efimkin.2017,Koepsell.2019,Fey.2020,glazov2020optical}. The role of electron-electron interactions on the properties of two-dimensional (2D) electron and hole gases has been extensively studied in GaAs~\cite{Chen.1990,Goldberg.1990,Zhu.2003}, AlAs~\cite{Gunawan.2006,DasSarma.2009} and recently also in graphene~\cite{Nomura.2006,Young.2012}. Typically the spin- and valley susceptibility of the system is found to be enhanced by interactions. Signatures of collective phenomena are particularly pronounced in 2D materials~\cite{Huang.2017,Tang.2020,Shimazaki.2020,Regan.2020} such as semiconducting transition metal dichalcogenide (TMDC) MoS$_2$ owing to strong Coulomb interaction~\cite{Chernikov.2014,Stier.2018}. Elegant control knobs, such as static electric-field effects have been used for tuning many-body interactions for the study of Fermi-polarons~\cite{Sidler.2016,Ravets.2018,Wang.2018,Smoleski.2019,Tan.2020,Liu.2020}. Specifically in the very low density regime, valley dichroism in the presence of a magnetic field generates a unique local spin and valley texture of excess charge carriers. Here, only few spin-polarized Landau levels (LLs) with low filling factors are involved in shaping the many-body environment. Therefore, the interaction of excitons immersed in the fermionic bath depends on those LLs. Quantum transport studies are intrinsically restricted to the high density regime where LLs with high filling factors are typically probed~\cite{Pisoni.2018,Movva.2017,Xu.2017,Gustafsson.2018,Larentis.2018,Fallahazad.2016,Lin.2019,Li.2020} whereas optical measurements facilitate access to the entire density regime~\cite{Smoleski.2019,Liu.2020}. The large variation of reported $g$-factors for excitons in MoS$_2$~\cite{Stier.2016,Mitioglu.2016,Cadiz.2017,Goryca.2019,Jadczak.2020} and other 2D materials~\cite{Aivazian.2015,Srivastava.2015,MacNeill.2015,Li.2014,Wang.2016,Stier.2016,Stier.2016b,Smoleski.2019,Lyons.2019,Goryca.2019,Liu.2020,Wang.2020} with arbitrary carrier concentration further highlights the necessity for studying valley Zeeman shifts with tunable and controlled carrier densities in 2D semiconductors.

\begin{figure}
	\scalebox{\figurescale}{\includegraphics[width=1\linewidth]{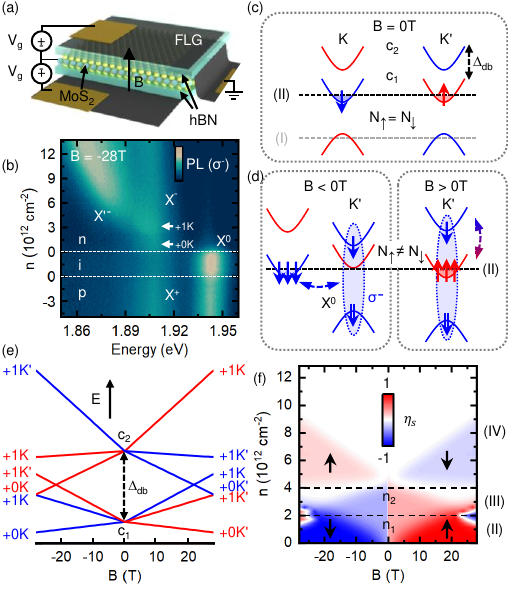}}
	\renewcommand{\figurename}{FIG.|}
	\caption{\label{fig1}
		(a) Schematic of the dual-gate monolayer MoS$_{2}$ vdW device in Faraday geometry. A monolayer MoS$_2$ is sandwiched between hexagonal boron nitride (hBN) and contacted with thin graphite (FLG) electrodes.
		(b) Carrier concentration dependent low-temperature ($T = \SI{5}{\kelvin}$) $\mu$-PL at $B = \SI{-28}{\tesla}$. Dashed lines show the region of charge neutrality ($n \sim \SI{0}{\per\centi\meter\squared}$). Arrows for $B = \SI{-28}{\tesla}$ highlight the $+0K$ and $+1K$ LLs in the $X^-$.	
		(c) Spin and valley exciton band structure at the $K$ and $K^{\prime}$ points of monolayer MoS$_{2}$ for restored time-reversal symmetry ($B = \SI{0}{\tesla}$).
		(d) A finite magnetic field ($B \neq \SI{0}{\tesla}$) breaks time-reversal symmetry inducing an electron spin imbalance ($N_{\uparrow} \neq N_{\downarrow}$).
		For positive magnetic field, the dipolar-spin interaction is of \textit{intra}valley-type, while for negative magnetic field of \textit{inter}valley-type.
		(e) Calculated LL fan diagram.
		(f) Calculated degree of spin-polarization $\eta_s$ showing the density and magnetic field dependent spin and valley texture in the conduction bands.
		}
\end{figure}

In this Letter, we fully address the variation of $g$-factors in the literature which is a direct consequence of the many-body interaction with the fermionic bath. We control the carrier concentration $n$ in a dual-gate field-effect device (Fig.~\ref{fig1}(a)) and study the magneto-optical response in high magnetic fields up to $B = \SI{28}{\tesla}$. Simultaneous control of $n$ and $B$ allows us to prepare a unique global spin texture originating from the quantization of excess carriers in fully spin-polarized LLs. We study the magnetic field dependence of neutral and charged excitons (Fermi-polarons) which encode the evolution of the total magnetic energy including carrier spin, valley magnetic moment (Berry phase) and cyclotron phenomena arising from quantization of electrons and/or holes into discrete LLs~\cite{Aivazian.2015,Srivastava.2015,MacNeill.2015,Li.2014,Wang.2016,Smoleski.2019,Liu.2020,Wang.2020}. From our measurements, we directly observe that shape and magnitude of the valley Zeeman shift $\Delta E_{VZ}$ of excitons very sensitively depend on the spin and valley texture. Our results suggest that the interaction of the exciton with the Fermi-bath at low densities is driven by dipolar spin-interaction which markedly differs from previous observations that have not taken into account the unique LL quantization in 2D TMDCs~\cite{Sidler.2016,Ravets.2018,Wang.2018,Smoleski.2019,Tan.2020,Liu.2020}.

We excite the sample with unpolarized light at $\lambda = \SI{514}{\nano\meter}$ and an excitation power of $\SI{30}{\micro\watt}$ and detect $\sigma^-$-polarized PL at $T = \SI{5}{\kelvin}$. Due to the robust optical selection rules in monolayer TMDCs, we only probe the emission from excitonic recombination in the $K^{\prime}$ valley for positive and negative polarities of the $B$-field~\cite{Xiao.2012}. Figure~\ref{fig1}(b) shows a carrier concentration dependent false colour PL map recorded at $B = \SI{-28}{\tesla}$. We symmetrically tune the voltage applied to the top and bottom gates, which effectively counteracts Fermi-level pinning due to the reduction of band tail states~\cite{Volmer.2020,Klein.2019}. This allows us to observe the $X^{+}$ transition in the p-charged regime I for the first time in MoS$_{2}$. The data reveal narrow emission lines of the $X^{0}$ due to the hBN encapsulation~\cite{Wierzbowski.2017,Cadiz.2017,Florian.2018} and LL oscillations in the~\textit{intra}valley $X^{-}$ at higher magnetic field. Importantly, another PL feature, labelled $X^{\prime-}$, appears red shifted from $X^{-}$ at $n_2 \sim 4 \cdot 10^{12} \SI{}{\per\centi\meter\squared}$. This peak quickly gains oscillator strength with increasing $n$ to dominate the spectrum at the highest value of $n$ studied. $X^{\prime-}$ is commonly observed for n-doping in MoS$_{2}$~\cite{Pisoni.2018,Roch.2020} and WSe$_{2}$~\cite{Barbone.2018}. It has recently been attributed to a Mahan-like exciton~\cite{Roch.2020}, or an exciton-plasmon-like excitation~\cite{VanTuan.2017}, but its precise origin is not yet fully understood. We repeated the measurement for various static magnetic fields ranging from $\SI{-28}{\tesla}$ to $\SI{28}{\tesla}$.

Figure~\ref{fig1}(c) depicts the spin and valley band sequence of monolayer MoS$_{2}$ in an excitonic picture. Electron-hole exchange leads to an optically dark alignment of the spin-orbit split conduction bands $c_1$ and $c_2$ ($\Delta_{db}\sim\SI{14}{\milli\electronvolt}$), which has been theoretically proposed~\cite{Deilmann.2017} and experimentally verified~\cite{Pisoni.2018,Robert.2020}. Throughout this Letter, we describe all observed effects on the basis of this picture. For $B = 0$ and $E_F$ situated slightly above the lower conduction band minima ($c_1$), an equal number of spin-up ($\uparrow$) and spin-down ($\downarrow$) electrons occupy the bands at $K$ and $K^{\prime}$ ($N_{\uparrow} = N_{\downarrow}$) resulting in zero net spin-polarization (Fig.~\ref{fig1}(c)). However, an applied magnetic field breaks time-reversal symmetry, shifts the conduction band minima, and the resident electrons condense into LLs resulting in an occupation imbalance between the valleys. A direct consequence is the emergence of a spin-polarized Fermi sea ($N_{\uparrow} \neq N_{\downarrow}$) in either $K$ or $K^{\prime}$ for positive and negative magnetic fields (Fig.~\ref{fig1}(d)). Therefore, the resulting spin-valley texture of the electrons depends on the carrier concentration ($E_F$) \textit{and} the direction and magnitude of the applied magnetic field as visualized for monolayer MoS$_2$ in Fig.~\ref{fig1}(f). For other monolayer TMDCs, this depends on the details of the respective band structure. In general, the global degree of spin-polarization $\eta_s = (N_{\uparrow}-N_{\downarrow})/(N_{\uparrow}+N_{\downarrow})$ enters different regimes ranging from $|\eta_s| = \pm 1$ at low carrier concentration and high magnetic fields (regime II), through intermediate densities with $|\eta_s| < 1$ (regime III) to the absence of $\eta_s$ at highest $n$ where $E_F$ resides well above the minimum of the upper conduction band $c_2$ (regime IV). For our consideration, we model only the relevant LLs in the conduction bands with a finite inhomogeneous broadening ($\SI{4}{\milli\electronvolt}$) and determine the number of spin-${\uparrow}$ and spin-${\downarrow}$ electrons from integrating over the corresponding density of states (see Fig.~\ref{fig1}(e) and further details in the Supplemental Material~\cite{Supplementary_Material}).

\begin{figure}
	\scalebox{\figurescale}{\includegraphics[width=1\linewidth]{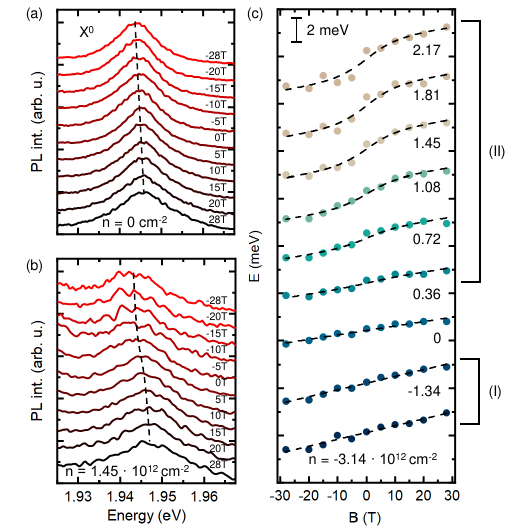}}
	\renewcommand{\figurename}{FIG.|}
	\caption{\label{fig2}
		(a) and (b) Normalized $\mu$-PL spectra of $X^{0}$ for applied magnetic fields at $n = \SI{0}{\per\centi\meter\squared}$ ($n = 1.45 \cdot 10^{12} \SI{}{\per\centi\meter\squared}$) from $\SI{-28}{\tesla}$ to $\SI{28}{\tesla}$. The solid lines are a guide to the eye.
		(c) Carrier density dependent valley Zeeman shift $\Delta E_{VZ}$ of the $X^{0}$ in the electron and hole charged regime.
		}
\end{figure}

An applied magnetic field lifts the $K/K^{\prime}$ valley degeneracy by shifting time-reversed pairs of states in opposite directions in accord with the Zeeman energy $-\mu_B \cdot B$~\cite{Xu.2014}. This effect will shift the exciton energy when the magnetic moment of conduction and valence bands are not equal, $\Delta E_{VZ} = -(\mu^c-\mu^v)\cdot B=\frac{1}{2}g \mu_B B$. Figures~\ref{fig2}(a) and \ref{fig2}(b) show the magnetic field dependent $X^0$ PL close to charge neutrality ($n \sim \SI{0}{\per\centi\meter\squared}$) and for low electron densities ($n \sim 1.45 \cdot 10^{12} \SI{}{\per\centi\meter\squared}$). While $\Delta E_{VZ}$ is completely linear at charge neutrality, strikingly it becomes nonlinear with electron doping. The observation is summarized in Fig.~\ref{fig2}(c), where the $B$-field dependent peak positions for a sequence of charge concentrations are shown.

At charge neutrality, we measure a $g$-factor of $X^{0}$, $g_{X^0} = -1.27 \pm 0.09$. This value is consistent with recent reports in high quality and hBN encapsulated MoS$_{2}$~\cite{Cadiz.2017,Jadczak.2020} but contrasts earlier work on non-encapsulated samples~\cite{Stier.2016,Mitioglu.2016}. Our measured $g$-factor appears to be unusually small compared to other TMDCs where a linear valley Zeeman shift for $X^{0}$ near $g \sim -4$ is reported~\cite{Li.2014,MacNeill.2015,Aivazian.2015,Srivastava.2015,Wang.2015,Stier.2016,Stier.2018,Goryca.2019}. In general, the total exciton magnetic moment has contributions from an orbital and a spin component \textbf{$\mu=\mu^{orb}+\mu^s$}, where $\mu^{orb}=-\frac{e}{2m_e}l_z$ and $\mu^s-\frac{eg_e}{2m_e}s_z$. Due to the hybridization of the $d_{x^2-y^2}$ and $d_{xy}$ orbitals in the valence band, the Bloch electrons at $K/K^{\prime}$ provide an angular momentum of $l_z=\pm 2 \hbar$ such that an estimate for the valley Zeeman splitting in a simple, non-interacting model is $\mu=-4 \mu_B$.~\cite{Srivastava.2015} Such $g$-factors are reproduced well by recent first-principle studies including excitonic effects~\cite{Deilmann.2020,Wozniak.2020}. However, as pointed out in Ref.~\cite{Wang.2015} and demonstrated in the Supplemental Material~\cite{Supplementary_Material}, the decomposition of $\mu^{orb}$ in a contribution from \textit{inter}-cellular hopping, captured by a lattice model, and atom-like \textit{intra}-cellular corrections is a delicate problem. Localized orbitals that do not allow for atom-like dipole transitions (e.g. $d_{z^2}$- and $d_{x^2-y^2}$/$d_{xy}$-orbitals in MoS$_2$) result in a vanishing \textit{intra}-cellular correction rendering $\mu^{orb} =0$. Breaking of the electron-hole symmetry is expected due to transitions from each band to energetically higher and lower lying bands resulting in a finite \textit{inter}-cellular contribution. However, even the most sophisticated theoretical approaches to date ignore the many-body nature of the electron-hole pair which we demonstrate here. We further note that the exciton $g$-factor results from couplings of energetically remote bands~\cite{Jovanov.2011}, which may be perturbed by free carriers as well as impurities~\cite{Klein.2019}. 

\begin{figure*}
	\scalebox{\figurescale}{\includegraphics[width=1\linewidth]{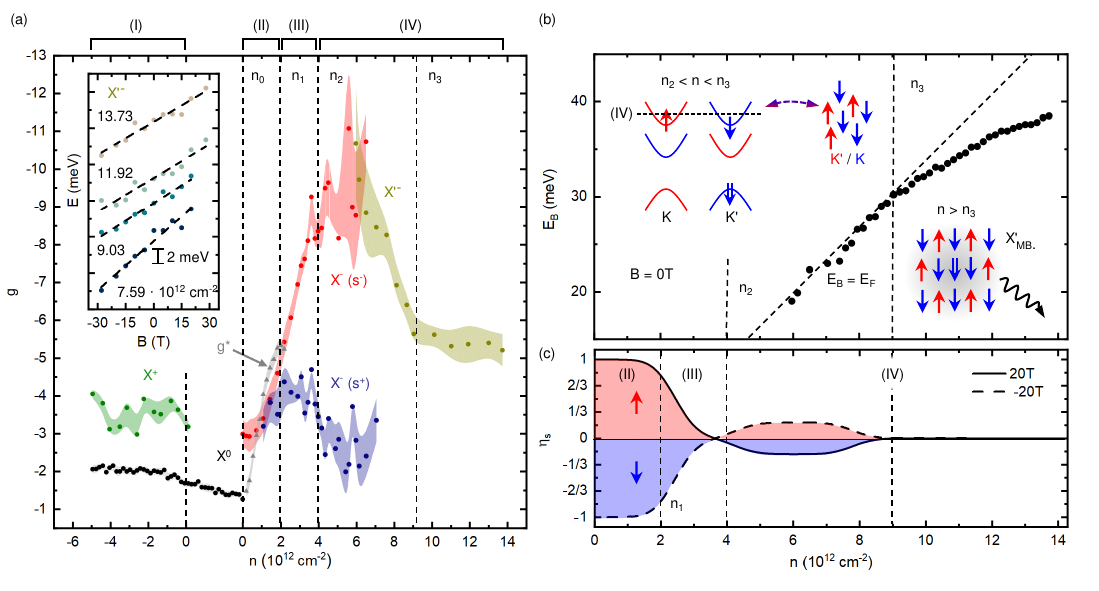}}
	\renewcommand{\figurename}{FIG.|}
	\caption{\label{fig3}
		(a) Carrier density dependent $g$-factors of $X^{0}$ (black and gray data), the negatively (red and blue data) and positively (green data) charged excitons $X^{-}$ and $X^{+}$, and the $X^{\prime-}$ feature (yellow data). 
		(b) Carrier density dependence of the $X^{\prime-}$ binding energy $E_B$. For $n_{2} < n < n_{3}$ $E_B$ is linear with $E_F$ while it becomes sub-linear for $n > n_{3}$. For the highest carrier densities (regime IV), the quasi-particle excitation is strongly interacting and dissolving into a many-body state $X_{MB.}^{\prime}$.
		(c) Electron concentration dependence of the degree of global spin-polarization $\eta_s$ for $B = \SI{-20}{\tesla}$ and $\SI{20}{\tesla}$.
		}
\end{figure*}

The nonlinear valley Zeeman shift in the n-charging regime directly correlates with $\eta_s$ and is a direct consequence of a dipolar spin-interaction of the exciton with the local spin-polarized fermionic bath. An enhancement in the context of conduction band filling has been observed in WSe$_2$ at high densities~\cite{Wang.2018}. However, our reported results clearly show that the interaction is driven by few electrons occupying fully spin-polarized LLs. Similar physics can be found for electrons interacting with local magnetic moments in dilute magnetic semiconductors~\cite{Furdyna.1988}. In our case, the effective magnetic moment originates from a magnetic field induced net spin-polarization of excess carriers as depicted schematically in Fig~\ref{fig1}(d). Qualitatively, this can be pictured as a paramagnetic spin ordering effect of conduction band electrons where the MoS$_{2}$ effectively undergoes a \textit{ferri}magnetic-to-\textit{ferro}magnetic phase transition~\cite{Lin.2019}. When a sufficiently large magnetic field is applied, the valley degeneracy is lifted, which leads to a repopulation of electrons into the lowest conduction band LL. As such a net spin imbalance, $S_Z(B)=-1/2\cdot B_J(x)$, between the valleys occurs at higher $B$. Here, the Brillouin function~\cite{Furdyna.1988} is $B_J(x) = \frac{2s_z+1}{2s_z} \coth{\bigg(\frac{2s_z+1}{2s_z}x\bigg)} - \frac{1}{2s_z} \coth{\bigg(\frac{1}{2s_z}x\bigg)}$ where $s_z = 1/2$ is the electron spin in $c_1$ and $x = g \mu_B s_z B / k_B T$. The net magnetization is then $M\sim \Delta N \cdot S_z(B)$ with the total number of polarized electron spins $\Delta N = |N_{\downarrow} - N_{\uparrow}|$. Therefore, the total valley Zeeman $g$-factor for the exciton in regime II can be expressed as the addition of spin and valley net magnetization effects as

\begin{equation}
\label{eq:valley_zeeman}
    g^*(n,B) = g_{X^0} + g_m(n)\cdot S_z(B)
\end{equation}

such that the nonlinear valley Zeeman shift is $\Delta E^{X^0}_{VZ}(n,B)= 1/2 g^*(n,B) \mu_B B$. As shown by the fits to the data in Fig.~\ref{fig2}(c), this simple model captures the nonlinear valley Zeeman shift surprisingly well. Note that we do not observe a nonlinearity in the p-charging regime I. This may be expected as interaction effects are more prominent in the conduction band~\cite{Pisoni.2018}.

Finally, we summarize the valley susceptibility of all features in our accessible density range in Fig.~\ref{fig3}(a).

\textit{Regime I.}---The $X^0$ $g$-factor monotonically increases from $\sim -1.27$ at charge neutrality to $-2.1$ at the highest hole density while the $X^{+}$ $g$-factor exhibits a value of $-(3.52 \pm 0.35)$ suggesting that the extra hole adds $\sim 1.5 \mu_B$ to the valley susceptibility of the exciton. Most strikingly, we find that $g^*$ monotonically increases from the value observed close to charge neutrality to the trion $g$-factor, showing the direct dependence of the exciton $g$-factor on the carrier density and strongly pointing towards the exciton having the character of a spin-polarized magnetic polaron at finite electron density.

\textit{Regime II.}---Introducing electrons to the system and using Eq.~\ref{eq:valley_zeeman} to obtain $g^*(n,B)$, the magnetic moment of $X^{0}$ (see Fig.~\ref{fig2}(c)) increases monotonically with carrier concentration, peaking at $g^* \sim -5.5$ for a density of $n_1 \sim 2 \cdot 10^{12} \SI{}{\per\centi\meter\squared}$. This confirms our expectation that for $n < n_1$, the increased magnetic susceptibility of $X^0$ arises from the density of spin-polarized electrons in the lower $K^{\prime}$ valley (see Fig.~\ref{fig1}(f)) and the magnitude of $g^*(n,B)$ is highly sensitive to the local electron concentration. This conclusion may explain the large variation in the literature of reported $g$-factors for excitons in MoS$_2$~\cite{Stier.2016,Mitioglu.2016,Cadiz.2017,Goryca.2019,Jadczak.2020}. Crucially, $g^*(n,B)$ smoothly approaches the $g$-factor observed for $X^{-}$ at higher electron concentrations, indicating that both states have a similar spin and valley structure. The $X^{0}$ interacts with the spin-polarized Fermi sea while the strength of the interaction is tuned by the Fermi energy, very similar to the Kondo problem of an isolated impurity spin interacting with a spin-polarized Fermi sea~\cite{Latta.2011}. For high densities, the spin structure of $X^{0}$ increasingly resembles that of $X^{-}$. Indeed, at the carrier density of $n_1$, $X^{0}$ sees on average $\sim 0.1$ electrons within its wave function.

\textit{Regime III.}---Increasing the electron concentration from moderate ($n_{1}$) to high ($n_{2}$) densities, we find an asymmetry of the $X^{-}$ valley Zeeman shift in positive ($X^{-}(s^{+})$) and negative ($X^{-}(s^{-})$) magnetic fields (see Fig. S5 in the Supplemental Material~\cite{Supplementary_Material}). We attribute this to LL occupation differences in the $K$/$K^\prime$ valley in the lower conduction bands $c_1$ (see Fig.~\ref{fig1}(e)). The $g$-factor for negative $B$ increases to $\sim -10$. Similar to $X^{0}$, $X^{-}$ also interacts with electrons in the Fermi sea. However, unlike $X^{0}$ the local spin valley texture admits electrons residing in both valleys since $E_F$ is located in the lower conduction bands (regime III). This picture is also supported by the spin texture, shown in Fig.~\ref{fig3}(c).

\textit{Regime IV.}---For $n > n_2$, the PL is quickly dominated by the feature $X^{\prime-}$ which emerges precisely at the carrier density when $E_F$ shifts into the upper conduction bands $c_2$ ($n_2 \sim 4 \cdot 10^{12} \SI{}{\per\centi\meter\squared}$ in our device)~\cite{Pisoni.2018}. We find that $\Delta E_{VZ}$ of $X^{\prime-}$ is symmetric and linear in $B$ for all accessible densities (see inset Fig.~\ref{fig3}(a)). The $g$-factor of $X^{\prime-}$ equals $g_{X^-(s^-)}$ at $n_2$, which suggests a qualitatively similar magnetic moment. Indeed, a bound complex with the excess electron occupying the $\nu = +0K$ LL in the $c_2$ is such a configuration (see inset Fig.~\ref{fig3}(b)). This is further substantiated by the very strong valley polarization of this feature (see Fig. S6 and S7 in the Supplemental Material~\cite{Supplementary_Material}). Both, $\eta_s$ and the $g$-factor of $X^{\prime-}$ simultaneously diminish above $n_3$ with $\eta_s \sim 0$ and a minimum $g$-factor of $\sim -5$. Strikingly, at exactly this concentration, a change in the dependence of the $X^{\prime-}$ binding energy $E_B = E(X^-) - E(X^{\prime-})$ occurs from a linear to a sub-linear dependence on $E_F$ (see Fig.~\ref{fig3}(b)). For $n > n_3$, all valleys and electron spin species are available since $E_F$ is situated well within $c_2$. The electron concentration approaches the Mott density~\cite{Steinhoff.2017} where strong many-body effects start to dominate the interaction of $X^{\prime-}$ and electrons of all spins forming a strongly dressed many-body state (see inset Fig.~\ref{fig3}(b)). Here the precise local spin structure becomes less relevant, and mean field-"like" theories become applicable. A detailed description of this state would call for dedicated many-body calculations of the spin susceptibility.

Our results show that all excitons in 2D materials are many-body correlated states that have a magneto-optical response that is sensitive to the local carrier density and related spin and valley textures. We explain the large variation of $g$-factors observed in the literature as arising from lack of control of local doping. The findings of our study represent an important step towards studying and engineering many-body related phases and novel interaction phenomena in atomically thin materials. \\

%
%
Supported by Deutsche Forschungsgemeinschaft (DFG) through the TUM International Graduate School of Science and Engineering (IGSSE) and the German Excellence Cluster-MCQST and e-conversion. We gratefully acknowledge financial support by the PhD program ExQM of the Elite Network of Bavaria, financial support from the European Union’s Horizon 2020 research and innovation programme under grant agreement No. 820423 (S2QUIP) the German Federal Ministry of Education and Research via the funding program Photonics Research Germany (contracts number 13N14846) and the Bavarian Academy of Sciences and Humanities. J.K. acknowledges support by the Alexander von Humboldt foundation. M.F., A.S. and F.J. were supported by the Deutsche Forschungsgemeinschaft (DFG) within RTG 2247 and through a grant for CPU time at the HLRN (Berlin/G\"ottingen). J.J.F and A.H. acknowledge support from the Technical University of Munich - Institute for Advanced Study, funded by the German Excellence Initiative and the European Union FP7 under grant agreement 291763 and the German Excellence Strategy Munich Center for Quantum Science and Technology (MCQST). K.W. and T.T. acknowledge support from the Elemental Strategy Initiative conducted by the MEXT, Japan, Grant Number JPMXP0112101001, JSPS KAKENHI Grant Number JP20H00354 and the CREST(JPMJCR15F3), JST. The work has been partially supported by the EC Graphene Flagship project (no. 604391), by the ANR projects ANR-17-CE24-0030 and ANR-19-CE09-0026. We further thank Scott Crooker and Mark Goerbig for insightful and stimulating discussions. \\

%
%

\bibliographystyle{naturemag}
\bibliography{full}

\end{document}


\title{Supplemental Material - Controlling exciton many-body states by the electric-field effect in monolayer MoS$_2$}
%
\author{J.~Klein}\email{jpklein@mit.edu}
\affiliation{Walter Schottky Institut and Physik Department, Technische Universit\"at M\"unchen, Am Coulombwall 4, 85748 Garching, Germany}
\affiliation{Department of Materials Science and Engineering, Massachusetts Institute of Technology, Cambridge, Massachusetts 02139, USA}
\author{A.~H\"otger}
\affiliation{Walter Schottky Institut and Physik Department, Technische Universit\"at M\"unchen, Am Coulombwall 4, 85748 Garching, Germany}
\author{M.~Florian}
\affiliation{Institut für Theoretische Physik, Universität Bremen, P.O. Box 330 440, 28334 Bremen, Germany}
\author{A.~Steinhoff}
\affiliation{Institut für Theoretische Physik, Universität Bremen, P.O. Box 330 440, 28334 Bremen, Germany}
\author{A.~Delhomme}
\affiliation{Universit\'e Grenoble Alpes, INSA Toulouse, Univ. Toulouse Paul Sabatier, EMFL, CNRS, LNCMI, 38000 Grenoble, France.}
\author{T.~Taniguchi}
\affiliation{Research Center for Functional Materials, National Institute for Materials Science, 1-1 Namiki, Tsukuba 305-0044, Japan
}
\author{K.~Watanabe}
\affiliation{Research Center for Functional Materials, National Institute for Materials Science, 1-1 Namiki, Tsukuba 305-0044, Japan
}
\author{F.~Jahnke}
\affiliation{Institut für Theoretische Physik, Universität Bremen, P.O. Box 330 440, 28334 Bremen, Germany}
\author{A.~W.~Holleitner}
\affiliation{Walter Schottky Institut and Physik Department, Technische Universit\"at M\"unchen, Am Coulombwall 4, 85748 Garching, Germany}
\author{M.~Potemski}
\affiliation{Universit\'e Grenoble Alpes, INSA Toulouse, Univ. Toulouse Paul Sabatier, EMFL, CNRS, LNCMI, 38000 Grenoble, France.}
\author{C.~Faugeras}
\affiliation{Universit\'e Grenoble Alpes, INSA Toulouse, Univ. Toulouse Paul Sabatier, EMFL, CNRS, LNCMI, 38000 Grenoble, France.}
\author{J.~J.~Finley}\email{finley@wsi.tum.de}
\affiliation{Walter Schottky Institut and Physik Department, Technische Universit\"at M\"unchen, Am Coulombwall 4, 85748 Garching, Germany}
\author{A.~V.~Stier}\email{andreas.stier@wsi.tum.de}
\affiliation{Walter Schottky Institut and Physik Department, Technische Universit\"at M\"unchen, Am Coulombwall 4, 85748 Garching, Germany}
%

%
\date{\today}
%

%
\maketitle
%
%

\tableofcontents

\newpage

\section{Field-effect device for carrier density control in monolayer MoS$_{2}$}

\begin{figure}[ht]
\scalebox{\figurescale}{\includegraphics[width=0.6\linewidth]{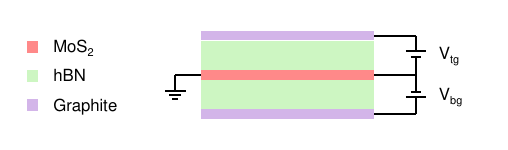}}
\renewcommand{\figurename}{Figure S}
\caption{\label{SIfig1}
Schematic of the field-effect van der Waals device. Monolayer MoS$_{2}$ is encapsulated between hBN. The device is a dual-gate device where the carrier density is controlled with a top- and bottom-gate $V_{tg}$ and $V_{bg}$. The same voltage with same polarity is applied to both gates for enhancing the gating effect.
}
\end{figure}

We make use of field-effect devices to control the carrier density in monolayer MoS$_{2}$.~\cite{CastellanosGomez.2014} Figure~S\ref{SIfig1} shows the two device geometries used in this manuscript. In our device, monolayer MoS$_{2}$ is encapsulated in hBN. We use the hBN for two main reasons: (i) Encapsulation reduces inhomogeneous linewidth broadening of excitons~\cite{Wierzbowski.2017} and (ii) as gate dielectric that withstands high breakdown fields,~\cite{Dean.2010} thus preventing leakage currents. We use few-layer graphite as the gate-electrode and to directly contact the MoS$_2$. The heterstacks are assembled by the dry viscoelastic transfer technique iteratively stacking the individual layers using PDMS stamping.~\cite{CastellanosGomez.2014} The dual-gate device with top and bottom gates as shown in Fig.~S\ref{SIfig1}. We apply equal gate voltages with the same polarity to the gates $V_{bg}=V_{tg}$ with respect to the monolayer MoS$_{2}$ for controlling the carrier density. In this device, we are also able to tune from the n- into the p-doped regime. The observation that only the n-doped regime is accessible with a single-gate is common in the literature for monolayer MoS$_{2}$.~\cite{Robert.2018,Roch.2019} It is likely that the dual-gate device allows to access the p-doped regime since it overcomes Fermi-level pinning effects by the symmetric device geometry. Furthermore, dual gates allow larger applied effective fields for tuning the carrier density. We determine the carrier density by using a simple plate capacitor model where the device capacitance is $C = \epsilon_0 \epsilon_{hBN}/d$ with the dielectric constant of multilayer hBN $\epsilon_{hBN} = 2.5$~\cite{Dean.2010,Kim.2012,Hunt.2013,Laturia.2018} and the hBN layer thickness $d = \SI{14}{\nano\meter}$ which is determined by atomic force microscopy (AFM). Since top and bottom hBN thickness are very similar for the dual-gate device we relate, the carrier density with the gate voltage through $n = C (V_{tg} + V_{bg}) /e = 2CV /e$.

\newpage

%
%

\section{Transfer characteristics in high magnetic fields}

%
\begin{figure}[!ht]
\scalebox{\figurescale}{\includegraphics[width=1\linewidth]{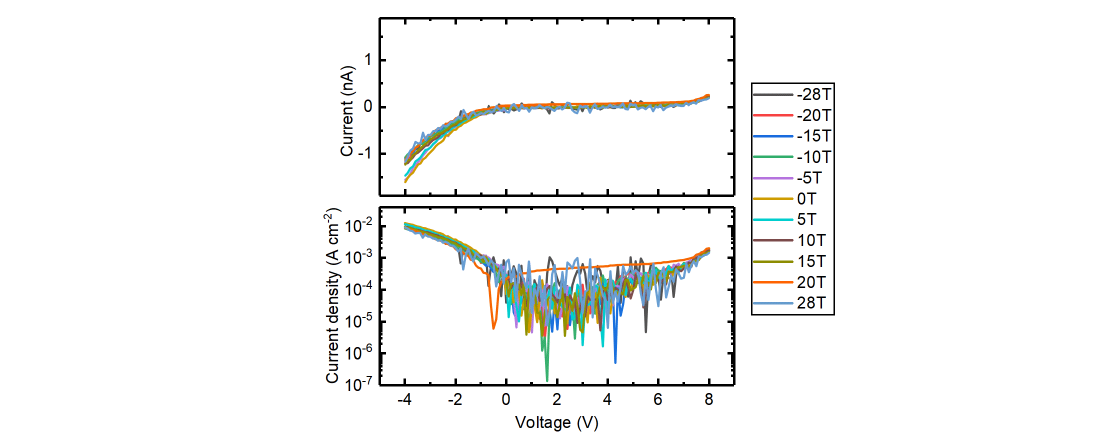}}
\renewcommand{\figurename}{Figure S}
\caption{\label{SIfig2}
Transfer characteristics of the dual-gate device. Top panels: I-V curves for all magnetic fields applied. Bottom panels: Current density as a function of applied bias voltage ($V=V_{bg}=V_{tg}$). The monolayer MoS$_{2}$ is excited with $\SI{30}{\micro\watt}$ at $\SI{514}{\nano\meter}$ with a laser spot diameter at the sample of $\sim \SI{4}{\micro\meter}$.
}
\end{figure}

For the gate-dependent magneto-photoluminescence measurements, we apply a magnetic field and vary the gate voltage in steps of $\SI{100}{\milli\volt}$ while collecting PL spectra for every gate voltage step. We perform the same gate biasing sequence for every magnetic field, thus ensuring that the voltage sweeps at different magnetic fields are directly comparable. We first apply a static magnetic field and then we tune the bias voltage from max. $V_{+}$ to max. $V_{-}$. Typical current voltage characteristics are presented in Fig.~S\ref{SIfig2}(a) and (b). We apply equal voltage to top- and bottom-gate ($V=V_{bg}=V_{tg}$). From the transfer characteristics of both devices, we find that our biasing scheme is highly reproducible for all magnetic fields applied in the experiment. The reproducibility is due to the graphite contacts to the MoS$_{2}$ which is known for low contact resistance and small Schottky barrier heights.~\cite{Cui.2015,Allain.2015} The leakage currents are in the noise floor for most of the range and negligible leakage currents of $<\SI{1.5}{\nano\ampere}$, that correspond to current densities of $< 10^{-2}\SI{}{\ampere\per\centi\meter\squared}$, at the highest bias voltages. The data are collected for a laser excitation power of $\SI{30}{\micro\watt}$ with a laser excitation energy of $\SI{2.41}{\electronvolt}$.

\newpage

\section{Carrier density dependent magneto-photoluminescence of monolayer MoS$_{2}$}

\begin{figure}[!ht]
\scalebox{\figurescale}{\includegraphics[width=1\linewidth]{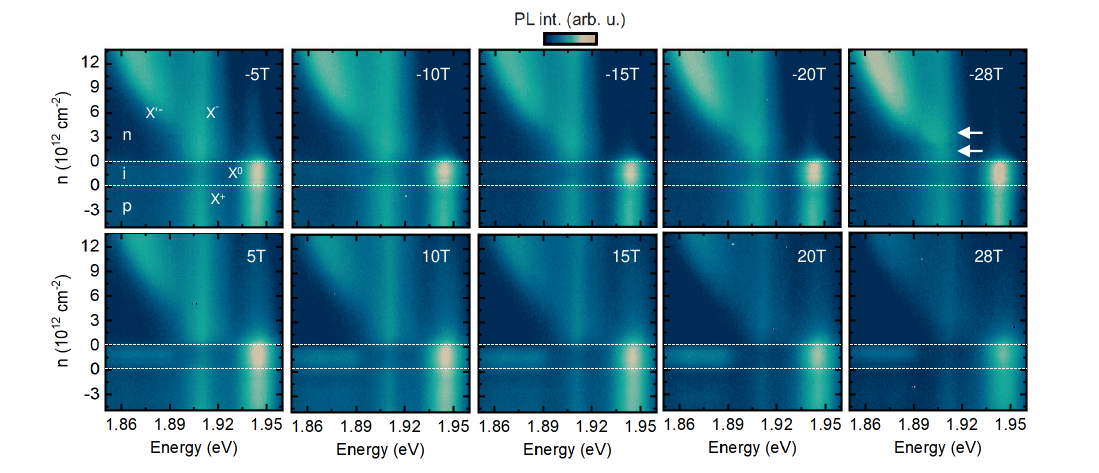}}
\renewcommand{\figurename}{Figure S}
\caption{
Carrier density dependent low-temperature ($T = \SI{5}{\kelvin}$) $\mu$-PL for static magnetic fields ranging from $\SI{-28}{\tesla}$ to $\SI{28}{\tesla}$. The $X^0$, $X^-$, $X^+$ and $X^{\prime-}$ PL features are marked.
}
\label{SIfig5}
\end{figure}

The $\sigma^-$ circularly polarized, charge carrier density dependent magneto-photoluminescence is shown in Fig.~S\ref{SIfig5}. The above described biasing sequence is used to maintain sample stability throughout the individual voltage sweeps for static magnetic fields ranging from $B = \SI{-28}{\tesla}$ to $B = \SI{28}{\tesla}$.

%
%

\newpage

\section{Quantized spin-valley texture}

The Zeeman shift of electrons in the spin-orbit split conduction band valleys in monolayer MoS$_2$ manifests from spin, the Berry phase and due to quantization of electrons in Landau levels in each valley. The shift for the lower conduction band $c_1$ is given by

\begin{equation}
    E_{c_1} = \tau_s s_z 2 \mu_B B + \tau_i m_e \mu_B B + \nu \frac{ \hbar e B}{m_e}~,
    \label{Eq:Ec1}
\end{equation}

while the shift of the upper conduction band $c_2$ is

\begin{equation}
    E_{c_2} = \Delta_{db} + \tau_s s_z 2 \mu_B B + \tau_i m_e \mu_B B + \nu \frac{ \hbar e B}{m_e}~.
    \label{Eq:Ec2}
\end{equation}

Here, the valley and spin indices are $\tau_s = \pm 1$ ($K=+1K$ and $K^{\prime}=-1K$) and $\tau_i = \pm 1$ ($+1$ spin-$\uparrow$ and $-1$ spin-$\downarrow$). We used an electron mass of $m_e = 0.44 m_0$. Moreover, $\nu$ is the filling factor of the LL and $\Delta_{db}$ the energy splitting between $c_1$ and $c_2$ for no magnetic field applied. We model two LLs for each valley with $\nu = +0$ and $\nu = +1$ in $K^{\prime}$.~\cite{Rose.2013} After quantifying the energy shift of each LL of each spin in every valley, we can further calculate the density of states (DOS) as a function of the Fermi level $E_F$ (applied gate voltage). From this quantity, we can then infer the number of electrons populating the spin-$\downarrow$ and spin-$\uparrow$ LLs to deduce the degree of spin polarization for a given $E_F$. We model each LL by using a Gaussian function

\begin{equation}
     DOS_{LL} = \frac{e |B|}{h}\frac{1}{\sigma \sqrt{2 \pi}} \exp{-\frac{(E-E_{LL})^2}{2\sigma^2}}~,  
\end{equation}

with $eB/h$ as the degeneracy per unit area and the energetic position of the LL $E_{LL}$ as defined in Eq.~\ref{Eq:Ec1} and~\ref{Eq:Ec2} and a FWHM of each LL of $\Gamma = 2 \sqrt{2 ln(2)} \sigma = \SI{4}{\milli\electronvolt}$ accounting for the experimentally observed inhomogeneous broadening. The total DOS for all spin-$\uparrow$ electrons is given through

\begin{equation}
     DOS^{\uparrow}_{LL} = \sum_{i=0}^{\nu=1} DOS^{\uparrow, K^{\prime}}_{LL} + \sum_{i=1}^{\nu=2} DOS^{\uparrow, K}_{LL}~,
\end{equation}

and the DOS for all spin-$\downarrow$ electrons are given by

\begin{equation}
     DOS^{\downarrow}_{LL} = \sum_{i=1}^{\nu=2} DOS^{\downarrow, K}_{LL} + \sum_{i=0}^{\nu=1} DOS^{\downarrow, K^{\prime}}_{LL}~.
\end{equation}

By integrating the DOS to $E_F$ we obtain the number of electrons populating each LL with the total number of electrons with spin-$\uparrow$

\begin{equation}
     N_{\uparrow} = \int_0^{E_F} DOS^{\uparrow}_{LL} dE
\end{equation}

and spin-$\downarrow$

\begin{equation}
     N_{\downarrow} = \int_0^{E_F} DOS^{\downarrow}_{LL} dE~.
\end{equation}

We can now compute the global degree of spin polarization all magnetic fields and $E_F$

\begin{equation}
    \eta_s(B,E_F) = \frac{N_{\downarrow} - N_{\uparrow}}{N_{\downarrow} + N_{\uparrow}}~.
\end{equation}

\newpage

\section{Magnetic moment of Bloch electrons: lattice Hamiltonian vs. atomic contributions}

The magnetic moment is comprised of a contribution due to the orbital motion of a Bloch electron and a contribution due to the electron spin. The z-component of the orbital magnetic moment is given by:
\begin{equation}
\mu^{\textrm{orb},n}_{z\boldsymbol{k}}=-\frac{e}{2 m_e}\langle\Phi^{n}_{\boldsymbol{k}}|\hat{l}_z|\Phi^{n}_{\boldsymbol{k}}\rangle
=-\frac{e}{2 m_e}\langle\Phi^{n}_{\boldsymbol{k}}|\hat{x}\hat{p}_y-\hat{y}\hat{p}_x|\Phi^{n}_{\boldsymbol{k}}\rangle.
\label{m_z_start}
\end{equation}
Consider the general expression $\langle\Phi^{n}_{\boldsymbol{k}}|\hat{x}_i\hat{p}_j|\Phi^{n}_{\boldsymbol{k}}\rangle$ and insert a complete set of Bloch states:
\begin{equation}
\langle\Phi^{n}_{\boldsymbol{k}}|\hat{x}_i\hat{p}_j|\Phi^{n}_{\boldsymbol{k}}\rangle=
\sum_{n'\boldsymbol{k}'}\langle\Phi^{n}_{\boldsymbol{k}}|\hat{x}_i|\Phi^{n'}_{\boldsymbol{k}'}\rangle\langle\Phi^{n'}_{\boldsymbol{k}'}|\hat{p}_j|\Phi^{n}_{\boldsymbol{k}}\rangle.
\label{xipj_completeness}
\end{equation}
The momentum matrix elements are diagonal in $\boldsymbol{k}$ due to translational invariance:
\begin{equation}
\langle\Phi^{n'}_{\boldsymbol{k}'}|\hat{p}_j|\Phi^{n}_{\boldsymbol{k}}\rangle=\langle\Phi^{n'}_{\boldsymbol{k}}|\hat{p}_j|\Phi^{n}_{\boldsymbol{k}}\rangle\delta_{\boldsymbol{k},\boldsymbol{k}'}.
\label{momentum_diag}
\end{equation}
The position matrix element can be transformed using the Schrödinger equation of Bloch states,
\begin{equation}
H\big|\Phi^{n}_{\boldsymbol{k}}\big>=\varepsilon_{\boldsymbol{k}}^{n}\big|\Phi^{n}_{\boldsymbol{k}}\big>,
\label{sgl_bloch}
\end{equation}
and the commutator relation~\cite{Gajdo.2006}
\begin{equation}
\frac{1}{m_e}\boldsymbol{\hat{p}}=\frac{i}{\hbar}\big[{H},{\boldsymbol{\hat{r}}}\big],
\label{H_r_comm}
\end{equation}
which holds in case of a local one-electron potential. It is still valid in the presence of spin-orbit interaction, as long as the latter can be approximately treated as an on-site potential.
Using Eqs. (\ref{momentum_diag}), (\ref{sgl_bloch}) and (\ref{H_r_comm}), we find: 
\begin{equation}
\langle\Phi^{n}_{\boldsymbol{k}}|\hat{x}_i\hat{p}_j|\Phi^{n}_{\boldsymbol{k}}\rangle=
\sum_{n'}\frac{\hbar}{i m_e}\langle\Phi^{n}_{\boldsymbol{k}}|\hat{p}_i|\Phi^{n'}_{\boldsymbol{k}}\rangle\frac{1}{\varepsilon_{\boldsymbol{k}}^{n}-\varepsilon_{\boldsymbol{k}}^{n'}}\langle\Phi^{n'}_{\boldsymbol{k}}|\hat{p}_j|\Phi^{n}_{\boldsymbol{k}}\rangle.
\label{xipj_transform}
\end{equation}
The crystal wave functions can be constructed as a linear combination of localized orbitals in the following way such that they fulfill Bloch's theorem:
\begin{equation}
\big|\Phi^{n}_{\boldsymbol{k}}\big\rangle=\sum_{\alpha}c^{n}_{\alpha}(\boldsymbol{k})\big|\boldsymbol{k}\alpha\big\rangle,\quad\big|\boldsymbol{k}\alpha\big\rangle=\frac{1}{\sqrt{N}}\sum_{\boldsymbol{R}}e^{i\boldsymbol{k}\cdot\boldsymbol{R}}\big|\boldsymbol{R}\alpha\big\rangle
\label{TB_ansatz}
\end{equation}
with the orthonormality relations
\begin{equation}
\big\langle \boldsymbol{R}\alpha \big|\boldsymbol{R}'\alpha'\big\rangle=\delta_{\boldsymbol{R}\boldsymbol{R}'}\delta_{\alpha\alpha'}
\label{orth_wann} 
\end{equation}
and
\begin{equation}
\big\langle\boldsymbol{k}\alpha\big|\boldsymbol{k}'\alpha'\big\rangle=\delta_{\boldsymbol{k}\boldsymbol{k}'}\delta_{\alpha\alpha'},
\label{orth_bloch}
\end{equation}
where $N$ is the number of lattice sites. We can formulate the crystal (lattice) Hamiltonian in terms of the localized orbitals:
\begin{equation}
H=\sum_{\boldsymbol{R}\boldsymbol{R}'\alpha\alpha'}t^{\alpha\alpha'}_{\boldsymbol{R}\boldsymbol{R}'}\big|\boldsymbol{R}\alpha\big\rangle\big\langle \boldsymbol{R}'\alpha' \big|.
\label{H_lat}
\end{equation}
Inserting the ansatz (\ref{TB_ansatz}) into Eq.~(\ref{xipj_transform}), we obtain:
\begin{equation}
\langle\Phi^{n}_{\boldsymbol{k}}|\hat{x}_i\hat{p}_j|\Phi^{n}_{\boldsymbol{k}}\rangle=
\sum_{n'}\frac{\hbar}{i m_e}\frac{1}{\varepsilon_{\boldsymbol{k}}^{n}-\varepsilon_{\boldsymbol{k}}^{n'}}
\sum_{\alpha\alpha'}(c^{n}_{\alpha}(\boldsymbol{k}))^*c^{n'}_{\alpha'}(\boldsymbol{k})
\langle\boldsymbol{k}\alpha|\hat{p}_i|\boldsymbol{k}\alpha'\rangle
\sum_{\alpha\alpha'}(c^{n'}_{\alpha'}(\boldsymbol{k}))^*c^{n}_{\alpha}(\boldsymbol{k})
\langle\boldsymbol{k}\alpha'|\hat{p}_j|\boldsymbol{k}\alpha\rangle.
\label{xipj_final}
\end{equation}
Following~\cite{Tomczak.2009}, we analyze the momentum matrix element by using the commutator relation (\ref{H_r_comm}) again, transforming the momentum states according to (\ref{TB_ansatz}) and inserting a complete set of position states:
\begin{equation}
\langle\boldsymbol{k}\alpha'|\hat{p}_j|\boldsymbol{k}\alpha\rangle=-\frac{i m_e}{\hbar}\frac{1}{N}\sum_{\boldsymbol{R},\boldsymbol{R}'}e^{i\boldsymbol{k}\cdot(\boldsymbol{R}-\boldsymbol{R}')}
\int \boldsymbol{dr}\big[ \langle\boldsymbol{R}'\alpha'|\hat{r}_j|\boldsymbol{r}\rangle\langle\boldsymbol{r}|H|\boldsymbol{R}\alpha\rangle
-\langle\boldsymbol{R}'\alpha'|H|\boldsymbol{r}\rangle\langle\boldsymbol{r}|\hat{r}_j|\boldsymbol{R}\alpha\rangle\big].
\label{p_transform}
\end{equation}
One has to distinguish between the continuous space variable $\boldsymbol{r}$ and the discrete unit-cell label $\boldsymbol{R}$. The position operator acts as $\hat{r}_j|\boldsymbol{r}\rangle=r_j|\boldsymbol{r}\rangle$. The momentum matrix element contains contributions that can be directly related to the discrete lattice as well as contributions that arise due to the spatial extension of orbitals. We separate these contributions by shifting $\boldsymbol{r}\rightarrow\boldsymbol{r}+\boldsymbol{R}'$ in the first term and $\boldsymbol{r}\rightarrow\boldsymbol{r}+\boldsymbol{R} $ in the second term:
\begin{equation}
\begin{split}
\langle\boldsymbol{k}\alpha'|\hat{p}_j|\boldsymbol{k}\alpha\rangle=&-\frac{i m_e}{\hbar}\frac{1}{N}\sum_{\boldsymbol{R},\boldsymbol{R}'}e^{i\boldsymbol{k}\cdot(\boldsymbol{R}-\boldsymbol{R}')}
\int\boldsymbol{dr}  \\
&\big[\langle\boldsymbol{R}'\alpha'|r_j+R'_j|\boldsymbol{r}+\boldsymbol{R}'\rangle\langle\boldsymbol{r}+\boldsymbol{R}'|H|\boldsymbol{R}\alpha\rangle \\
&-\langle\boldsymbol{R}'\alpha'|H|\boldsymbol{r}+\boldsymbol{R}\rangle\langle\boldsymbol{r}+\boldsymbol{R}|r_j+R_j|\boldsymbol{R}\alpha\rangle\big] \\
=&-\frac{i m_e}{\hbar}\frac{1}{N}\sum_{\boldsymbol{R},\boldsymbol{R}'}e^{i\boldsymbol{k}\cdot(\boldsymbol{R}-\boldsymbol{R}')}(R'_j-R_j)\langle\boldsymbol{R}'\alpha'|H|\boldsymbol{R}\alpha\rangle \\
&-\frac{i m_e}{\hbar}\frac{1}{N}\sum_{\boldsymbol{R},\boldsymbol{R}'}e^{i\boldsymbol{k}\cdot(\boldsymbol{R}-\boldsymbol{R}')}
\int\boldsymbol{dr}r_j  \\
&\big[\langle\boldsymbol{R}'\alpha'|\boldsymbol{r}+\boldsymbol{R}'\rangle\langle\boldsymbol{r}+\boldsymbol{R}'|H|\boldsymbol{R}\alpha\rangle \\
&-\langle\boldsymbol{R}'\alpha'|H|\boldsymbol{r}+\boldsymbol{R}\rangle\langle\boldsymbol{r}+\boldsymbol{R}|\boldsymbol{R}\alpha\rangle\big],
\end{split}
\label{p_shift}
\end{equation}
where we made use of the completeness of position states again to derive the first term of the second line. This so-called Peierls contribution given by \textit{inter}-site hopping can be written as a generalized Fermi velocity:
\begin{equation}
\begin{split}
-\frac{i m_e}{\hbar}&\frac{1}{N}\sum_{\boldsymbol{R},\boldsymbol{R}'}e^{i\boldsymbol{k}\cdot(\boldsymbol{R}-\boldsymbol{R}')}(R'_j-R_j)\langle\boldsymbol{R}'\alpha'|H|\boldsymbol{R}\alpha\rangle \\
=\frac{m_e}{\hbar}&\frac{\partial}{\partial k_j}\frac{1}{N}\sum_{\boldsymbol{R},\boldsymbol{R}'}e^{i\boldsymbol{k}\cdot(\boldsymbol{R}-\boldsymbol{R}')}\langle\boldsymbol{R}'\alpha'|H|\boldsymbol{R}\alpha\rangle \\
=\frac{m_e}{\hbar}&\frac{\partial}{\partial k_j}\tilde{H}^{\alpha'\alpha}_{\boldsymbol{k}}.
\end{split}
\label{Fermi_velocity}
\end{equation}
It follows from the Schrödinger equation (\ref{sgl_bloch}) that the Hamiltonian $\tilde{H}^{\alpha\alpha'}_{\boldsymbol{k}}$ defines the tight-binding-like eigenvalue problem
\begin{equation}
\sum_{\alpha'}\tilde{H}^{\alpha\alpha'}_{\boldsymbol{k}}c^{n}_{\alpha'}(\boldsymbol{k})=\varepsilon_{\boldsymbol{k}}^{n}c^{n}_{\alpha}(\boldsymbol{k}).
\label{TB_eig}
\end{equation}
The second term in Eq.~(\ref{p_shift}) contains continuum contributions due to the finite extension of the electron wave functions. It can be written as
\begin{equation}
\begin{split}-\frac{i m_e}{\hbar}\frac{1}{N}\sum_{\boldsymbol{R},\boldsymbol{R}'}e^{i\boldsymbol{k}\cdot(\boldsymbol{R}-\boldsymbol{R}')}
\int\boldsymbol{dr}r_j\sum_{\boldsymbol{R}'',\alpha''}&\big[\langle\boldsymbol{R}'\alpha'|\boldsymbol{r}+\boldsymbol{R}'\rangle\langle\boldsymbol{r}+\boldsymbol{R}'|\boldsymbol{R}''\alpha''\rangle\langle\boldsymbol{R}''\alpha'' |H|\boldsymbol{R}\alpha\rangle \\
&-\langle\boldsymbol{R}'\alpha'|H|\boldsymbol{R}''\alpha''\rangle\langle\boldsymbol{R}''\alpha''|\boldsymbol{r}+\boldsymbol{R}\rangle\langle\boldsymbol{r}+\boldsymbol{R}|\boldsymbol{R}\alpha\rangle\big] \\
=-\frac{i m_e}{\hbar}\frac{1}{N}\sum_{\boldsymbol{R},\boldsymbol{R}'}e^{i\boldsymbol{k}\cdot(\boldsymbol{R}-\boldsymbol{R}')}
\int\boldsymbol{dr}r_j\sum_{\boldsymbol{R}'',\alpha''}&\big[(\chi_{\boldsymbol{R}'\alpha'}(\boldsymbol{r}+\boldsymbol{R}'))^*\chi_{\boldsymbol{R}''\alpha''}(\boldsymbol{r}+\boldsymbol{R}')\langle\boldsymbol{R}''\alpha'' |H|\boldsymbol{R}\alpha\rangle \\
&-\langle\boldsymbol{R}'\alpha'|H|\boldsymbol{R}''\alpha''\rangle(\chi_{\boldsymbol{R}''\alpha''}(\boldsymbol{r}+\boldsymbol{R}))^*\chi_{\boldsymbol{R}\alpha}(\boldsymbol{r}+\boldsymbol{R})\big]
\end{split}
\label{p_second_term}
\end{equation}
with the wave functions $\chi_{\boldsymbol{R}\alpha}(\boldsymbol{r})=\langle \boldsymbol{r} | \boldsymbol{R}\alpha \rangle $. This term accounts on the one hand for all atomic or \textit{intra}-site $(\boldsymbol{R}=\boldsymbol{R}')$ processes and on the other hand for corrections to the \textit{inter}-site processes contained in the Peierls term. Hence, we can split the momentum matrix element into three contributions: 
\begin{equation}
\langle\boldsymbol{k}\alpha'|\hat{p}_j|\boldsymbol{k}\alpha\rangle=\frac{m_e}{\hbar}\frac{\partial}{\partial k_j}\tilde{H}^{\alpha'\alpha}_{\boldsymbol{k}}+p^{\alpha'\alpha}_{\boldsymbol{k},j}\big|_{\textrm{inter-site corr.}}+p^{\alpha'\alpha}_{\boldsymbol{k},j}\big|_{\textrm{intra-site corr.}}\,.
\label{momentum_contrib}
\end{equation}
In the limit of well-localized orbitals, the dominant correction is given by the \textit{intra}-site term \cite{Tomczak.2009}. It is obtained by using the lattice periodicity, $\chi_{\boldsymbol{R}\alpha}(\boldsymbol{r}+\boldsymbol{R})=\chi_{\boldsymbol{0}\alpha}(\boldsymbol{r}) $, to shift the origins of all wave functions to the same unit cell, which is labeled $\boldsymbol{0}$. We then identify those terms where the wave function arguments also lie within the same unit cell ($\boldsymbol{R}''=\boldsymbol{R}' $ in the first term, $\boldsymbol{R}''=\boldsymbol{R} $ in the second term):
 \begin{equation}
 \begin{split}
p^{\alpha'\alpha}_{\boldsymbol{k},j}\big|_{\textrm{intra-site corr.}}&=-\frac{i m_e}{\hbar}\frac{1}{N}\sum_{\boldsymbol{R},\boldsymbol{R}'}e^{i\boldsymbol{k}\cdot(\boldsymbol{R}-\boldsymbol{R}')}
\int\boldsymbol{dr}r_j\sum_{\alpha''} \\
&\big[(\chi_{\boldsymbol{0}\alpha'}(\boldsymbol{r}))^*\chi_{\boldsymbol{0}\alpha''}(\boldsymbol{r})\langle\boldsymbol{R}'\alpha'' |H|\boldsymbol{R}\alpha\rangle
-\langle\boldsymbol{R}'\alpha'|H|\boldsymbol{R}\alpha''\rangle(\chi_{\boldsymbol{0}\alpha''}(\boldsymbol{r}))^*\chi_{\boldsymbol{0}\alpha}(\boldsymbol{r})\big]\\
&=-\frac{i m_e}{\hbar}\int \boldsymbol{dr}r_j\sum_{\alpha''}
\big((\chi_{\boldsymbol{0}\alpha'}(\boldsymbol{r}))^* \chi_{\boldsymbol{0}\alpha''}(\boldsymbol{r}) \tilde{H}^{\alpha''\alpha}_{\boldsymbol{k}}  
-\tilde{H}^{\alpha'\alpha''}_{\boldsymbol{k}}(\chi_{\boldsymbol{0}\alpha''}(\boldsymbol{r}))^* \chi_{\boldsymbol{0}\alpha}(\boldsymbol{r}) 
\big)\\
&=-\frac{i m_e}{\hbar}\sum_{\alpha''}\big(r_j^{\alpha'\alpha''}\tilde{H}^{\alpha''\alpha}_{\boldsymbol{k}}-\tilde{H}^{\alpha'\alpha''}_{\boldsymbol{k}}r_j^{\alpha''\alpha} \big)
\end{split}
\label{intra_site}
\end{equation}
with the atom-like dipole matrix elements
 \begin{equation}
 \begin{split}
r_j^{\alpha\alpha'}=\int \boldsymbol{dr}r_j (\chi_{\boldsymbol{0}\alpha}(\boldsymbol{r}))^* \chi_{\boldsymbol{0}\alpha'}(\boldsymbol{r})
\end{split}
\label{atomic_dipole}
\end{equation}
forcing the usual optical selection rules $\Delta l = \pm 1,\Delta m =0,\pm 1 $. Note that the \textit{intra}-site term (\ref{intra_site}) needs to take into account the same set of localized orbitals that is used to set up the lattice Hamiltonian (\ref{H_lat}). The fact that the coupling to the magnetic field via the vector potential has to be two-fold in a tight-binding or lattice approach has also been discussed in Ref.~\cite{Wang.2015}. There are always contributions that can not be captured by the so-called Peierls substitution~\cite{Tomczak.2009} in the lattice Hamiltonian leading to the first term in (\ref{momentum_contrib}). If we nevertheless use the Peierls contribution alone, we end up with the following lattice formulation of the orbital magnetic moment:
\begin{equation}
\begin{split}
\mu^{\textrm{orb},n}_{z\boldsymbol{k}}\big|_{\textrm{lat}}&=-\frac{e}{2 m_e}\frac{\hbar}{i m_e}\big(\frac{m_e}{\hbar}\big)^2\sum_{n'}\frac{1}{\varepsilon_{\boldsymbol{k}}^{n}-\varepsilon_{\boldsymbol{k}}^{n'}}
\sum_{\alpha\alpha'}(c^{n}_{\alpha}(\boldsymbol{k}))^*c^{n'}_{\alpha'}(\boldsymbol{k})
\frac{\partial}{\partial k_x}\tilde{H}^{\alpha\alpha'}_{\boldsymbol{k}}
\sum_{\alpha\alpha'}(c^{n'}_{\alpha'}(\boldsymbol{k}))^*c^{n}_{\alpha}(\boldsymbol{k})
\frac{\partial}{\partial k_y}\tilde{H}^{\alpha'\alpha}_{\boldsymbol{k}} \\
&-(x\leftrightarrow y) \\
&=\frac{ie}{2\hbar}\sum_{n'}\frac{1}{\varepsilon_{\boldsymbol{k}}^{n}-\varepsilon_{\boldsymbol{k}}^{n'}}
\Big\{\sum_{\alpha\alpha'}(c^{n}_{\alpha}(\boldsymbol{k}))^*c^{n'}_{\alpha'}(\boldsymbol{k})
\frac{\partial}{\partial k_x}\tilde{H}^{\alpha\alpha'}_{\boldsymbol{k}}
\sum_{\alpha\alpha'}(c^{n'}_{\alpha'}(\boldsymbol{k}))^*c^{n}_{\alpha}(\boldsymbol{k})
\frac{\partial}{\partial k_y}\tilde{H}^{\alpha'\alpha}_{\boldsymbol{k}}-c.c.\Big\} \\
&=-\frac{e}{\hbar}\sum_{n'}\frac{1}{\varepsilon_{\boldsymbol{k}}^{n}-\varepsilon_{\boldsymbol{k}}^{n'}}
\textrm{Im}\,\Big\{\sum_{\alpha\alpha'}(c^{n}_{\alpha}(\boldsymbol{k}))^*c^{n'}_{\alpha'}(\boldsymbol{k})
\frac{\partial}{\partial k_x}\tilde{H}^{\alpha\alpha'}_{\boldsymbol{k}}
\sum_{\alpha\alpha'}(c^{n'}_{\alpha'}(\boldsymbol{k}))^*c^{n}_{\alpha}(\boldsymbol{k})
\frac{\partial}{\partial k_y}\tilde{H}^{\alpha'\alpha}_{\boldsymbol{k}}\Big\}.
\end{split}
\label{m_z_lat}
\end{equation}
In a simple two-band model, the structure of this expression leads to equal orbital magnetic moments for conduction and valence electrons. Corrections are expected due to transitions from each band, respectively, to energetically higher and lower bands. Moreover, if the two fundamental bands are composed of localized orbitals that do not allow for atom-like dipole transitions (e.g. $d_{z^2}$- and $d_{x^2-y^2}$/$d_{xy}$-orbitals in transition metal dichalcogenide monolayers), the \textit{intra}-site correction will vanish. The exciton g-factor, calculated directly from the magnetic moments of conduction and valence bands at the $\boldsymbol{K}$-point as $g_X=2(\mu^{c}_{z\boldsymbol{K}}-\mu^{v}_{z\boldsymbol{K}})/\mu_{\textrm{B}}$, will therefore be zero since the net spin of the exciton is zero. This is consistent with the $k\cdot p$-picture discussed in Ref.~\cite{Wang.2015}.

\newpage

\section{Valley Zeeman shift of $X^{-}$, $X^{+}$ and $X^{\prime-}$}

%
%
\begin{figure}[!ht]
\scalebox{\figurescale}{\includegraphics[width=1\linewidth]{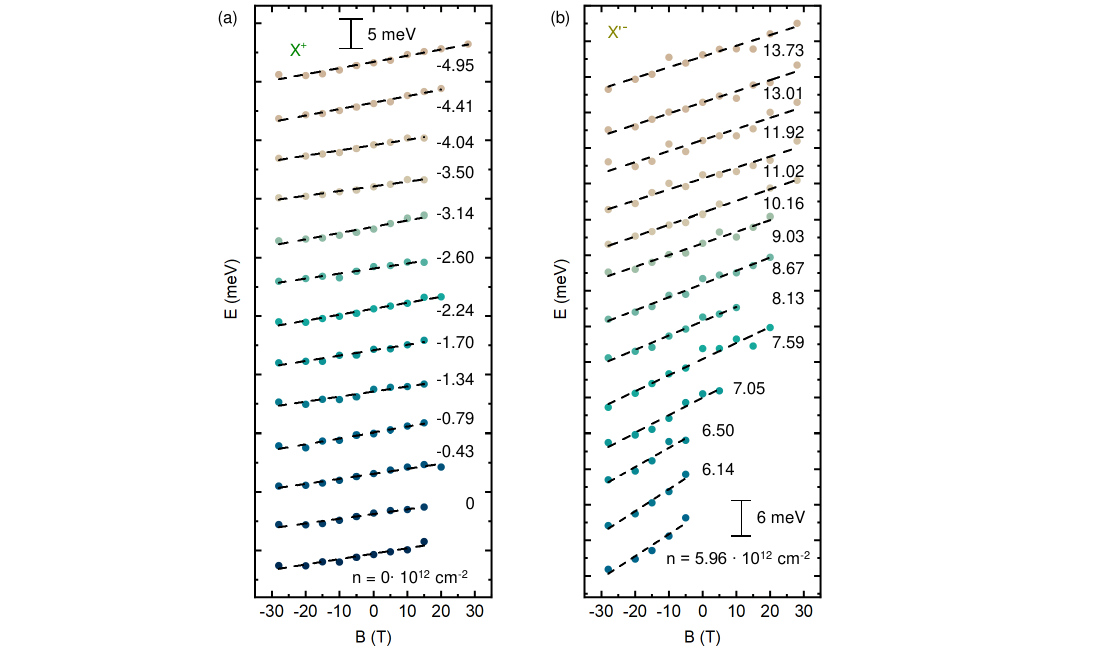}}
\renewcommand{\figurename}{Figure S}
\caption{
(a) Carrier density dependent Zeeman shift of the positively charged exciton $X^+$. (b) Carrier density dependent Zeeman shift of the strongly dressed high density feature $X^{\prime-}$.
}
\label{SIfig9}
\end{figure}

%
%
\begin{figure}[!ht]
\scalebox{\figurescale}{\includegraphics[width=1\linewidth]{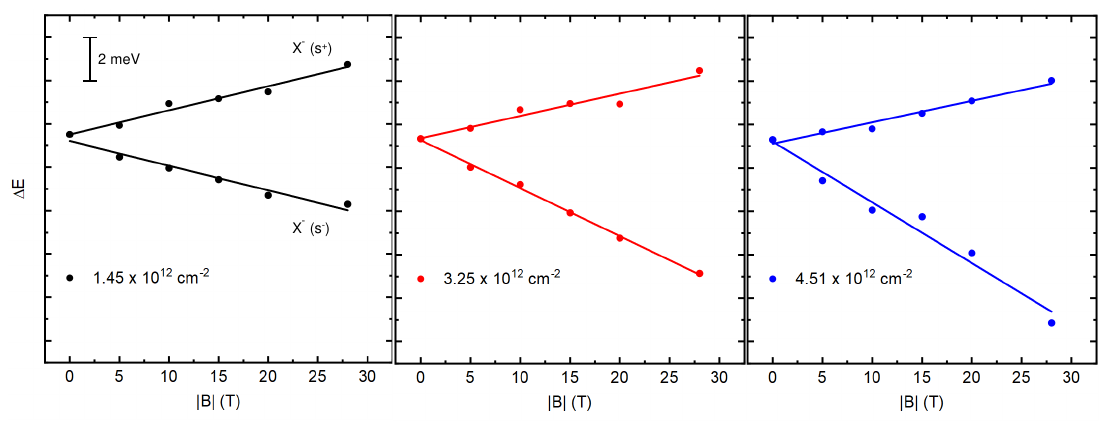}}
\renewcommand{\figurename}{Figure S}
\caption{
Valley Zeeman shift of the negatively charged exciton $X^-$ for positive (s+) and negative (s-) magnetic field direction for electron concentrations of $n = 1.45 \cdot 10^{12}\SI{}{\per\centi\meter\squared}$, $3.25 \cdot 10^{12}\SI{}{\per\centi\meter\squared}$ and $4.51 \cdot 10^{12}\SI{}{\per\centi\meter\squared}$.
}
\label{SIfig13}
\end{figure}

\newpage

\section{Magnetic field and density dependent valley dichroism}

%
%
\begin{figure}[!ht]
\scalebox{\figurescale}{\includegraphics[width=1\linewidth]{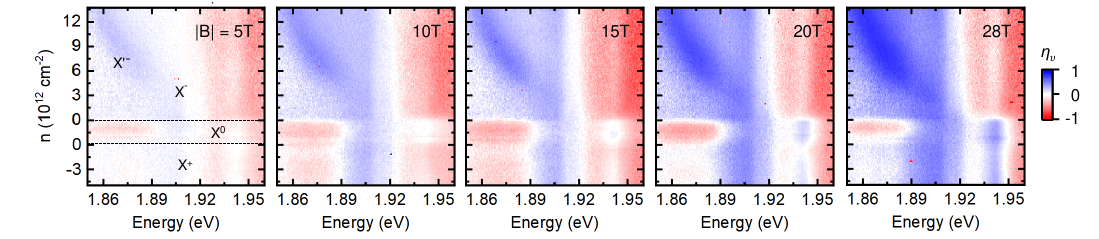}}
\renewcommand{\figurename}{Figure S}
\caption{
False color maps of the energy and carrier density dependent degree of valley polarization $\eta_v$ at $|B|=\SI{5}{\tesla}$, $\SI{10}{\tesla}$, $\SI{15}{\tesla}$, $\SI{20}{\tesla}$ and $\SI{28}{\tesla}$. Charge neutrality is highlighted with the dashed black line. The $X^-$ and $X^{\prime-}$ reveal enhanced degree of polarization for higher magnetic field.
}
\label{SIfig10}
\end{figure}

\newpage

%
%
\begin{figure}[!ht]
\scalebox{\figurescale}{\includegraphics[width=0.55\linewidth]{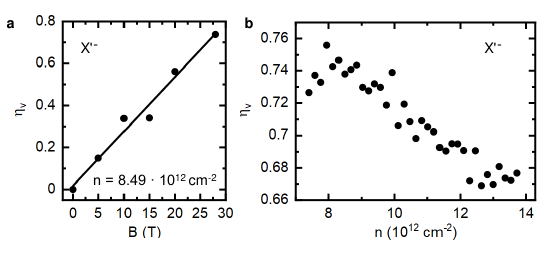}}
\renewcommand{\figurename}{Figure S}
\caption{
(a) Magnetic field dependence of the degree of valley polarization $\eta_v$ of the $X^{\prime-}$ for $n = 8.49 \cdot 10^{12} \SI{}{\per\centi\meter\squared}$. Solid line is a linear fit to the data.
(b) Electron concentration dependence of $\eta_v$ at $|B|=\SI{28}{\tesla}$.
}
\label{SIfig11}
\end{figure}

\newpage

%
%
\begin{figure}[!ht]
\scalebox{\figurescale}{\includegraphics[width=0.45\linewidth]{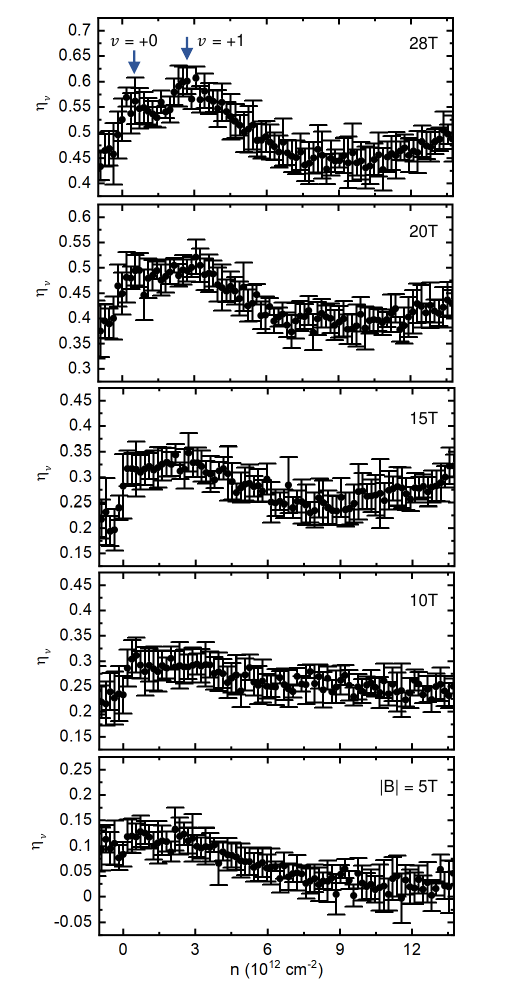}}
\renewcommand{\figurename}{Figure S}
\caption{
Electron density dependent degree of valley polarization of the \textit{intra}valley trion $X^-$ for magnetic fields ranging from $|B| = \SI{5}{\tesla}$ to $\SI{28}{\tesla}$. For $|B| = \SI{28}{\tesla}$, LLs with filling factors of $\nu = +0K$ and $\nu = +1K$ are highlighted by the two arrows, respectively. 
}
\label{SIfig12}
\end{figure}

\newpage

%
%
%
%
\bibliographystyle{apsrev}
\bibliography{full}